# Crystallite size dependence of morphotropic phase boundary in the multiferroic BiFeO3-PbTiO$_3$

Bastola Narayan, V. Kothai  and Rajeev Ranjan*

Department of Materials Engineering, Indian Institute of Science Bangalore-560012, India

## Abstract

The morphotropic phase boundary (MPB) in the magnetoelectric multiferroic BiFeO$_3$-PbTiO$_3$ is shown to be uniquely dependent on crystallite size. The system  exhibits the phenomenon of abnormal grain growth (AGG) during sintering with abrupt increase in the grain size from ~ 1 micron to ~ 10 microns. The large sized free crystallites (10 microns) exhibit pure tetragonal phase. Phase coexistence sets in after reducing the size to ~ 1 micron. These results prove that size, and not the internal stress, as the controlling mechanism for stabilizing phase coexistence in this system.

*rajeev@materials.iisc.ernet.in

The magnetoelectric multiferroic $BiFeO_3$ (BF) has been extensively investigated over the years in pure and modified forms because of its interesting ferroelectric and magnetic properties. BF exhibits rhombohedral (R3c) structure at room temperature and shows both ferroelectric and magnetic ordering in this phase [1]. The ferroelectric and the magnetic ordering set in below 850 C, and 350 C, respectively. Bulk $BiFeO_3$ however poses severe limitations such as extremely large coercive field, large leakage current and difficulty in obtaining pure perovskite phase by normal solid state sintering. Moreover, $BiFeO_3$ exhibits a long range spiral modulation of the canted antiferromagnetic spin which effectively cancels the macroscopic magnetization and thereby precludes the magneto-electric effect in the bulk state [2]. Attempts have been made to overcome these problems by chemical modifications. The solid solution series $BiFeO_3$-$PbTiO_3$ has been considerably investigated over the years [3-19]. This system exhibits very interesting features such as (i) formation of morphotropic phase boundary [3], (ii) unusually large tetragonality (~19 %) of the tetragonal phase [3-11], (iii) isostructural phase transformation [14, 17, 18] and (iv) pressure induced para-antiferromagnetic switching [20]. Since morphotropic phase boundary (MPB) systems are very interesting from the piezoelectric application point of view [22], the high Curie point of $BiFeO_3$-$PbTiO_3$ system has led to the anticipation that this can be used as high temperature piezoelectric material [7]. However, this potential has not been fully realized and there seems to be a lack of understanding with regard to the very nature of the reported MPB itself. Different research groups have reported different composition range of MPB for this system suggesting that the stability of the phases is highly sensitive to slight change in synthesis conditions. Fedulov et al [3] reported the MPB to lies in the range x=0.27 – 0.34 by sintering for 1 hour in the temperature range 800-1000 C. Woodward et al, on the other hand suggested the MPB range as 0.30<x<0.40 [6]. Bhatachajee et al reported the MPB to lies in the range 0.27<x<0.31 [7]. Zhu et al reported the widest range of MPB, i.e. 0.20<x<0.40 [12]. Siddaramanna et al have reported that a predominantly tetragonal phase could be stabilized in the composition x=0.20, which is generally reported to exhibit rhombohedral phase in the bulk state [21]. Very recently we have demonstrated that even experiments repeated under similar condition gave two different outcomes - sometimes the sintered pellet was found to be fragmented to powder, and on certain occasions it remained as a dense solid pellet after sintering. It was invariably noticed that the pellet consisted of R3c+ P4mm phase mixture whereas the powders were pure P4mm [19]. The nature of the phase stabilized at room temperature was found to be highly sensitive to small changes in the sintering temperature and time. For example, compositions such as 0.275, which is considered to be within the MPB range by majority of the research groups in the past [3, 7, 12], and even pure rhombohedral by Woodward et al [6], was shown to exhibit pure tetragonal phase when the sufficiently sintered. Another composition in the MPB range, x=0.30, exhibited phase coexistence when sintered at 980 C for 2 hours and showed pure tetragonal phase when the sintering duration was increased to 4 hours. When the sintering temperature was slightly reduced by 20 C, i.e. to 960 C it took 6 hours to yield a pure tetragonal phase [19]. Bell et al [10] and Comyn et al [11] have also reported thermal disintegration of pellets in certain composition range of this system. These studies although established that the phase formation behaviour is highly sensitive to specimen processing conditions such as sintering temperature, duration of sintering, cooling rates, etc. it could not establish as to what is the essential physical factor which determines if the phase formed was pure tetragonal or tetragonal + rhombohedral. Due to lack of direct evidence it was suggested that the rhombohedral phase is metastable in nature whose nuclei have certain life time at high temperatures [19]. If the specimen was cooled from the sintering temperature before the rhombohedral nuclei could disappear then the system would exhibit coexistence of rhombohedral and tetragonal phases. On the other hand, if the metastable rhombohedral nuclei disappear before cooling from the sintering temperature, the system would only show pure tetragonal phase. In this work we have identified that the most fundamental factor that determines the phase formation behaviour of the system is grain size. We found that this system exhibits the phenomenon of abnormal grain growth (AGG) during the sintering process. The grains were found to undergo abrupt grain growth from about 1 micron to ~ 10 microns. The 1 micron grains exhibit coexistence of tetragonal and rhombohedral phase whereas the abnormally grown grains (~10 microns) exhibit pure tetragonal phase. If during sintering AGG is allowed to happen then during cooling from the sintering temperature all the grains adopt tetragonal phase below the Curie point. In this situation the pellet undergoes spontaneous fragmentation because the grain boundaries lose their integrity due to the enormous stress associated with the large spontaneous strain (19 %) during the cubic to tetragonal transformation. On the other hand if the AGG is not allowed to happen during the sintering duration the pellet survives fragmentation because the small sized grains exhibit mixture of tetragonal and rhombohedral phases. The rhombohedral phase accommodates the large stress generated due to the formation of the tetragonal phase and shields the grain boundary from disintegration. This one-to-one correspondence between the occurrence of AGG and phase formation behaviour settles the long standing controversy with regard to the nature of MPB in this interesting multiferroic system.

Specimens of (1-$x$) BiFeO$_3$– ($x$) PbTiO$_3$ were prepared by conventional solid state route using analytical grade high purity powders of Bi$_2$O$_3$, PbO, Fe$_2$O$_3$ and TiO$_2$. The powders were mixed thoroughly for about 6 hours using Fritch Ball mill in a zirconia bowl with zirconia balls and acetone as the mixing media. The mixed powders were calcined at 800°C for 2 h. The calcined powder was pelletized under uniaxial loading and sintered in the range (depending on the composition) 950 - 1100 °C on alumina plate surrounded by the powders of the same composition and covered with an alumina crucible to minimize the volatilization of PbO and Bi$_2$O$_3$. X--ray powder diffraction patterns were collected using XPERTPro, PANalytical diffractometer using CuK$\alpha$ radiation. Rietveld refinement was carried out using the program FULLPROF [22].

Fig. 1 shows the x-ray powder diffraction pattern of x=0.29, a composition which is generally considered to be within the MPB range. Different pellets of this composition were sintered at 970 C for different durations to determine the time required for spontaneous disintegration for sintering at this temperature. It was found that the pellets survived disintegration up to 9 hours of sintering. On increasing the sintering duration to one more hour i.e. 10 hours of sintering, the pellet completely disintegrated to powder. The XRD patterns of the specimens sintered for 2 hours, 9 hours revealed coexistence of tetragonal and rhombohedral phases, whereas that of the 10 hours sintered specimen revealed pure tetragonal phase. The scanning electron microscope (SEM) images of the three differently sintered specimens are shown beneath their respective XRD patterns in Fig. 1. The grain sizes of the 2 hours and 9 hours sintered specimens were found to be in the range 0.5 – 1 micron while the crystallites of the 10 hours sintered specimen shows average size ~ 10 microns. This is a case of abrupt increase in the average grain size by one order of magnitude in between the 9[th] and the 10[th] hour of sintering. A similar scenario was found for another composition x=0.35. Woodward et al [6] and Zhu et [12] have reported phase coexistence for this composition whereas Fedulov et al [3], Bhattacharjee et al [9] and Comyn et al [20] have reported pure tetragonal phase. In our experiments we found the pellets of this composition to be spontaneously disintegrated after sintering at 1050 $^o$C for 6 hours. The x-ray powder diffraction of this powder exhibits pure tetragonal phase. The average grain size of the spontaneously disintegrated powder was found to be ~ 10 microns (Fig. 2). These two examples categorically prove that the formation of the tetragonal phase occurs in the abnormally grown grains. We have verified this to be true for other compositions of this series in the range 0.27<x<0.45. This also turns out to the composition range where confusion persists in the literature with regard to the stability of the phase(s) in this system.

Since the tetragonality (c/a-1) of the tetragonal phase is ~ 19 %, as the specimen is cooled below the Curie point the clamped grains in the dens ceramic specimen experience stress associated with the large tetragonal spontaneous strain. Bell et al [10] and Comyn et al [11] have suggested that the rhombohedral phase appears as a result of stress relief mechanism in the system. Though, in view of the fact that those specimens which exhibit phase coexistence (P4mm + R3c) remain as dens ceramic after sintering, and those without the rhombohedral phase disintegrates to powder, it appears plausible to suggest that rhombohedral phase indeed provides a stress relief mechanism, it still does not explain why this mechanism is not operative all the time. Our experiments and those reported by others [10, 11] have shown that given sufficient sintering temperature and time the system still prefer to stabilize pure tetragonal phase leading the pellets to fragment to powder. The question therefore arises as to whether the stress relief is at the origin of the two phase (MPB) state or, the appearance of the rhombohedral phase is purely due to smaller sized grains. Since, due to spontaneous disintegration of the pellet, the abnormally grown crystallites exhibiting pure tetragonal phase are unclamped and hence stress free, it was possible for us to examine the possibility of pure size effect by reducing the size of these big sized free grains. A small quantity of the powder whose SEM and XRD pattern is shown in the top left panel of Fig. 2 was thoroughly ground in a agate mortar pestle. To get rid of the stress induced ferroelastic changes, if any, incurred while grinding the powder, the ground powder was annealed at ~ 800 C for 6 hours. While this temperature is reasonably well above the Curie point (550 C) for this composition (x=0.35), is still not too high for the grains to coalesce. The bottom panel of Fig. 2 shows the SEM image and XRD patterns of the spontaneously disintegrated powder of x=0.35 and after its thorough grinding. Evidently the average size of the grains in the ground powder is drastically reduced to less than a micron. The XRD pattern of this powder shows rhombohedral peaks. Rietveld fitting of this pattern by rhombohedral + tetragonal phase coexistence model gave 29 percent of rhombohedral phase. The refined structural parameters are shown in Table I. Since, unlike in the pellets the grains are free, the formation of the rhombohedral phase cannot not be attributed to a stress relaxation mechanism. This proves that the formation of the rhombohedral phase is solely due to the smaller sizes grains. The size induced transformation of pure tetragonal state to MPB can be understood in terms of an equivalent compressive pressure resulting from specific surface energy of finite size crystal. For small crystallites faceting may be ignored and an average specific surface energy $\gamma_{av}$ may be considered. For a spherical grain of radius $r$, the capillary action of the surface energy $\gamma_{av}$ will bring about a net compressive stress $2\gamma_{av}/r$. It is this self-generated compressive stress which seems primarily responsible for partly bringing about a tetragonal to rhombohedral transformation. Very recently Comyn et al [20] have shown that an external hydrostatic pressure of mere 0.4 GPa can induce tetragonal to rhombohedral transformation in a composition x=0.30 of this series. The equivalence of the hydrostatic pressure and size reduction in bringing about the tetragonal to rhombohedral transformation seems to validate our view point that the compressive pressure induced by the enhanced surface energy in the smaller sized grains is the driving force which stabilizes the rhombohedral phase in this system.

In conclusion, we have established a one-to-one correlation between the stability of ferroelectric phase(s) and grain size in the multiferroic system $BiFeO_3$-$PbTiO_3$. The system was found to exhibit an unusually rare phenomenon of abnormal grain growth.

During this process the grain size of the specimen abruptly changes from ~1 micron to ~ 10 microns. If the synthesis conditions allow abnormal grain growth to happen the resulting large sized grains exhibit pure tetragonal phase. On the other hand if abnormal grain growth is not allowed to happen the average grain size remain ~ 1 micron and they exhibit coexistence of tetragonal and rhombohedral phases. It is further proved that partial stabilization of the rhombohedral phase is not a result of stress relief mechanism but is purely due to size driven tetragonal to rhombohedral phase transition. This study therefore resolves the five decade old controversy and provides explanation for the different MPB ranges reported for this system in the past.

**References**

1.  G. Catalan, and J. F. Scott, Adv. Mater. 21, 2463 (2009)
2.  P. Fischer, M. Polomska, I. Sosnowska, and M. Szymanski, J. Phys. C. Solid State Phys. 13, 1931 (1980)
3.  S. A. Fedulov, P. B. Ladyzhinskii, I. L. Pyatigorskaya and Yu. N. Venevetsev, Soviet. Phys. –Solid State **6**, 375 (1964)
4.  R. T. Smith, G. D. Achenbach, R. Gerson, and W. J. James, J. Appl. Phys. **39**, 70 (1968)
5.  V. V. S. Sai Sunder, A. Halliyal, and A. M. Umarji, J. Mater. Res. **10**, 1301 (1995)
6.  D. I. Woodward, I. M. Reaney, R. E. Eitel, and C. A. Randall, J. Appl. Phys. **94**, 3313 (2003)
7.  T. P. Comyn, T. Stevenson, and A. J. Bell, J. Phys. IV France 128, 13 (2005)
8.  J. Chen, X. R. Xing, G. R. Liu, J. H. Li and Y. T. Liu, Appl. Phys. Lett. **89**, 01914 (2006)
9.  S. Bhattacharjee, S. Tripathi and D. Pandey, Appl. Phys. Lett. **91**, 042903 (2007)
10. A. J Bell, A. X Levander. S. L. Turner, and T. P. Comyn, "Proceedings of 16th IEEE International Symposium on the Applications of Ferroelectrics", Pages: 406-409,(2007); DOI: 10.1109/ISAF.2007.4393280
11. T. P. Comyn, T. Stevenson, M. Al-Jawad, S. L. Turner, R. I. Smith, W. G. Marshall, A. J. Bell and R. Cywinski, Appl. Phys. Lett. 93, 232901 (2008)

12. W. −M. Zhu, H. −Y. Guo, and Z. G. Ye, Phys. Rev. B 78, 014401 (2008)

13. T. P. Comyn, T. Stevenson, M. Al-Jawad. S. L. Turner, R. I. Smith, A. J. Bell, and R. Cywinski, J. Appl. Phys. **105**, 094108 (2009)

14. R. Ranjan and A. Raju, Phys. Rev. B. **82**, 054119 (2010).

15. R.Ranjan, V.Kothai, Anatoliy Senyshyn, and Hans Boysen, J. Appl. Phys. **109**, 063522 (2011)

16. T. P. Comyn, T. Stevenson, M. Al-Jawad, G. Andre, A. J. Bell, and R. Cywinski, J. Mag. Mag. Mater. 323, 2533 (2011).

17. S.Bhattacharjee, K.Taji, C.Moriyoshi, Y.Kuroiwa and D.Pandey, Phys. Rev. B **84**, 104116 (2011)

18. V. Kothai, A.Senyshyn and R. Ranjan , J. Appl. Phys**,113**, 084102 (2013)

19. V. Kothai, R. Prasath, and R. Ranjan J. Appl. Phys. 114, 114102 (2013)

20. T. P. Comyn, T. Stevenson, M. Aj-Jawad, W. G. Marshall, R. I. Smith, J. H. Albilos, R. Cywinski, and A. J. Bell, J. Appl. Phys. 113, 183910 (2013)

21. A. Siddaramanna, V. Kothai, C. Srivastava, and R. Ranjan, J. Phys. D Appl. Phys. 47 045004 (2014)

22. Rodrigues-J. Carvajal, *FullPROF.A Rietveld Refinement and Pattern Matching Analysis Program* (Laboratories Leon Brillouin (CEA-CNRS), France, 2000).

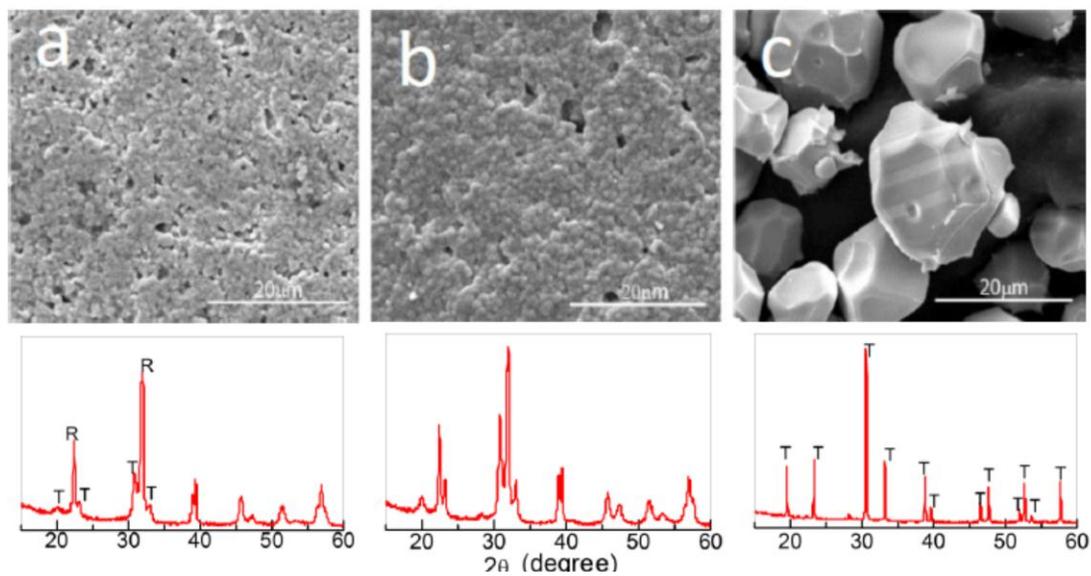

Fig.1 SEM micrographs of x=0.29 sintered at 970°C for (a) 2h, (b) 9h and (c) 10h. The XRD patterns of the the corresponding specimens are shown below.  R and T denote the rhombohedral and tetragonal peaks, respectively.

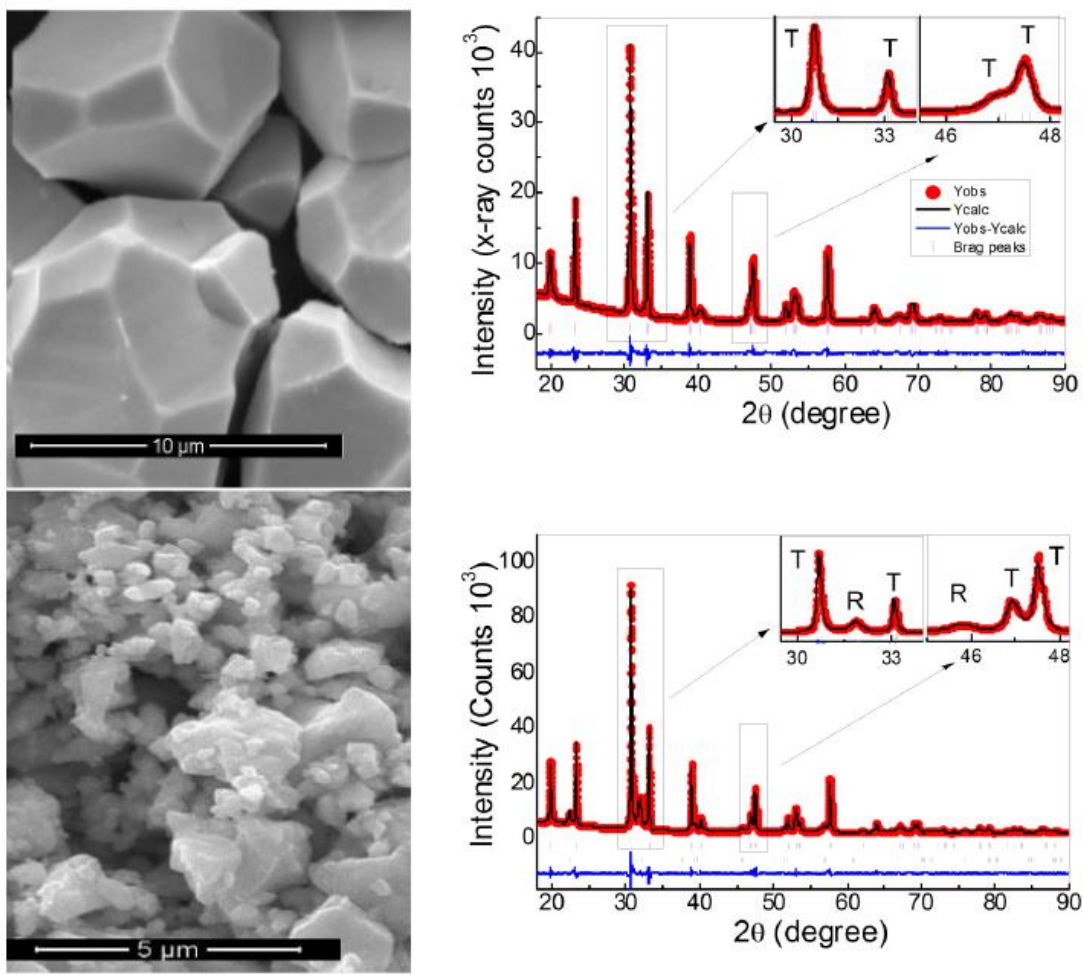

Fig.2 SEM image spontaneously disintegrated pellet of x=0.35 (top figure) and throughly ground disintegrated powder (bottom figure). The Rietveld fitted XRD pattern of the corresponding powder is shown on the right. R and T denote the rhombohedral and tetragonal peaks, respectively.

Table 1: Refined structural parameters and agreement factors of the thoroughly ground disintegrated powder of  0.65 BiFeO3-0.35PbTiO3 using tetragonal (*P4mm*) + Rhombohedral (R3c) phase coexistence model.

| | Space group: P4mm | | | | Space group: R3c | | | |
|---|---|---|---|---|---|---|---|---|
| Atoms | X | y | Z | B($\text{Å}^2$) | x | Y | z | B($\text{Å}^2$) |
| Bi/Pb | 0.000 | 0.000 | 0.000 | 1.918(3) | 0.000 | 0.000 | 0.2206(2) | 1.8(1) |
| Fe/Ti | 0.500 | 0.5000 | 0.5723(9) | 0.661(9) | 0.500 | 0.000 | -0.052(3) | 0.2(1) |
| O1 | 0.500 | 0.5000 | 0.1632(2) | 1.2(4) | 0.1325 | 0.8715 | 1/12 | 0.1(2) |
| O2 | 0.500 | 0.000 | 0.6697(1) | 0.2(1) | | | | |
| a= 3.8252(1) Å, c= 4.4845 (1) Å | | | | | a= 5.6118 (4)Å, b= 5.61118(4)Å, c= 13.6873(5) | | | |
| $R_{Bragg}$: 1.88, v= 65.618(3) $\text{Å}^3$ , %Phase = 71 (1) | | | | | $R_{Bragg}$: 2.16, v= 373.298(1) $\text{Å}^3$, %Phase = 29(1) | | | |
| $R_p$: 11.4,   $R_{wp}$: 11.9,   $R_{exp}$: 3.46, Chi$^2$:  11.7 | | | | | | | | |